\documentclass[two column,superscriptaddress,groupedaddress amssymb, amsmath,nobibnotes,nofootinbib, aps, showpacs, prb]{revtex4-1}
\usepackage[dvips]{graphicx}
\usepackage{graphicx}
\usepackage{color}

\begin{document}

\title{Multigap superconductivity in RbCa$_2$Fe$_4$As$_4$F$_2$ investigated using $\mu$SR measurements}
\author{D.T Adroja}
\email{devashibhai.adroja@stfc.ac.uk}
\affiliation{ISIS Facility, Rutherford Appleton Laboratory, Chilton, Didcot Oxon, OX11 0QX, United Kingdom} 
\affiliation{Highly Correlated Matter Research Group, Physics Department, University of Johannesburg, PO Box 524, Auckland Park 2006, South Africa}
\author{F. K. K. Kirschner}
\affiliation{Department of Physics, University of Oxford, Clarendon Laboratory, Parks Road, Oxford OX1 3PU, United Kingdom} 
\author{F. Lang} 
\affiliation{Department of Physics, University of Oxford, Clarendon Laboratory, Parks Road, Oxford OX1 3PU, United Kingdom} 
\author{M. Smidman}
\email{msmidman@zju.edu.cn}
\affiliation{Center for Correlated Matter and Department of Physics, Zhejiang University, Hangzhou 310058, China}
\author{A.D. Hillier}
\affiliation{ISIS Facility, Rutherford Appleton Laboratory, Chilton, Didcot Oxon, OX11 0QX, United Kingdom} 
\author{Zhi-Cheng Wang}
\affiliation{Department of Physics and State Key Lab of Silicon Materials, Zhejiang University, Hangzhou 310027, China}
\author{Guang-Han Cao}
\affiliation{Department of Physics and State Key Lab of Silicon Materials, Zhejiang University, Hangzhou 310027, China}
\author{G. B. G. Stenning}
\affiliation{ISIS Facility, Rutherford Appleton Laboratory, Chilton, Didcot Oxon, OX11 0QX, United Kingdom} 
\author{S. J. Blundell} 
\affiliation{Department of Physics, University of Oxford, Clarendon Laboratory, Parks Road, Oxford OX1 3PU, United Kingdom}

\date{\today}

\begin{abstract}

The superconducting properties of the recently discovered double Fe$_2$As$_2$ layered high-$T_c$ superconductor RbCa$_2$Fe$_4$As$_4$F$_2$ with $T_c\approx$ 30~K have been investigated using magnetization, heat capacity, transverse-field (TF) and zero-field (ZF) muon-spin rotation/relaxation ($\mu$SR) measurements. Our  low field magnetization measurements and heat capacity (C$_p$) reveal an onset of bulk superconductivity with $T_{\bf c}\sim$ 30.0(4) K. 
Furthermore, the heat capacity exhibits a jump at $T_{\bf c}$ of  $\Delta$C$_p$/$T_{\bf c}$=94.6 (mJ/mole-K$^2$) and no clear effect of applied magnetic fields was observed on C$_p$(T) up to 9 T between 2 K and 5 K. 
Our analysis of the TF-$\mu$SR results shows that the temperature dependence of  the magnetic penetration depth  is better described by a two-gap model, either isotropic  $s$+$s$-wave or $s$+$d$-wave than a single gap isotropic $s$-wave or $d$-wave model for the superconducting gap. The presence of two superconducting gaps in RbCa$_2$Fe$_4$As$_4$F$_2$ suggests a multiband nature of the superconductivity, which is consistent with the multigap superconductivity observed in  other Fe-based superconductors, including ACa$_2$Fe$_4$As$_4$F$_2$ (A=K and Cs).  Furthermore,  from our TF-$\mu$SR study we have estimated an in-plane penetration depth $\lambda_{\mathrm{ab}}$$(0)$ =231.5(3) nm, superconducting carrier density $n_s = 7.45 \times 10^{26}~ $m$^{-3}$, and carrier's effective-mass $m^*$ = 2.45\textit{m}$_{e}$. Our ZF $\mu$SR measurements do not reveal a clear sign of  time reversal symmetry breaking at $T_{\bf c}$, but the temperature dependent relaxation between 150 K and 1.2 K might indicate the presence of spin-fluctuations. The results of our present study have been compared with those reported for other Fe pnictide superconductors.
\end{abstract}

\pacs{74.70.Xa, 74.25.Op, 75.40.Cx}

\maketitle

\section{Introduction}

The discovery of high temperature superconductivity in fluorine-doped LaFeAsO (1111-family) with a transition temperature of $T_c$$\sim$26 K by Kamihara {\it et al.} has generated a considerable research interest world-wide to understand the nature of the superconductivity in this new class of compounds~\cite{Kamirara2006}. It was realized soon after the discovery that the $T_c$ of Fe-based superconductors can be increased up to 56 K as observed in Gd$_{0.8}$Th$_{0.2}$FeAsO~\cite {C.Wang2008}, Sr$_{0.5}$Sm$_{0.5}$FeAsF~\cite {G.Wu2009} and Ca$_{0.4}$Nd$_{0.6}$FeAsF~\cite{Cheng 2009}.  Until this discovery, high temperature superconductivity in cuprates, created the impression that only Cu-O planes are pivotal for understanding the mechanism of high temperature superconductivity~\cite{Anderson2013, Keimer2015}. Of course, the Fe-based superconductors do not contain Cu-O planes and some of the materials are even O free, for example, FeSe (11-family, $T_c$=8 K at ambient pressure and 46 K in applied pressure), LiFeAs (111-family, $T_c$ = 18 K), CaFeAs$_2$ (112-family, $T_c$ = 20~K) and ThFeAsN ($T_c$=30 K)~\cite {FeSe, LiFeAs, CaFeAs2, C.Wang2016}. 

Another highly investigated family of Fe-based superconductors is hole (i.e. K) and electron (i.e. Co, Ni, Rh and Pd) doped BaFe$_2$As$_2$ (122-family), which have a body centered tetragonal ThCr$_2$Si$_2$-type structure (I4/mmm), where the ubiquitous Fe$_2$As$_2$ layers of the Fe arsenide superconductors lie between the alkaline/alkaline earth atom layers shown in Fig.~1~\cite {122family1, 122family2, JPaglione2010,  XianhuiChen2014, MRotter2008}. Recently  superconductivity with $T_c$$\sim$35 K has been reported in Ca$A$Fe$_4$As$_4$ ($A$~=~K, Rb, Cs, 1144-family) \cite{1144Rep} and these materials consist of different arrangements of the layers along the c-axis also displayed in Fig.~1. In this structure, the alternating arrangement of the $A$ and Ca layers leads to two inequivalent As sites either side of the Fe sheets. The crystallographically inequivalent position of the Ca and A atoms changes the space group from I4/mmm (as for the 122-family) to P4/mmm.  Further, the different valence attraction from Ca$^{2+}$ and A$^{1+}$ layers to Fe$_2$As$_2$$^{1.5-}$ and the different ionic radii leads to different lengths of the As-Fe bonds, which was proposed to be an important parameter for controlling the $T_c$ of Fe-based superconductors~\cite{YMizuguchi2010}. Stoichiometric CaKFe$_4$As$_4$ is intrinsically near optimal hole doping~\cite{1144Rep,1144Prop} and does not exhibit a high temperature structural phase transition~\cite{W.R.Meier2016}. Similar to the optimally doped 122 compounds, probes of the gap structure and inelastic neutron scattering results strongly suggest the presence of a fully gapped $s_{\pm}$ state~\cite{CaKFe4As4ARPES,1144Gap1,1144Gap2,1144INS,1144NMR}. Further angle-resolved photoemission spectroscopy (ARPES) measurements of CaKFe$_4$As$_4$ report the presence of four superconducting gaps on different sheets of the Fermi surface~\cite {CaKFe4As4ARPES,1144Gap2}, which has been explained using a four-band $s_\pm$-wave Eliashberg theory emphasizing the important role of antiferromagnetic spin fluctuations~\cite {G.A.Ummarino2016}. 

\par
Very recently Zhi-Cheng Wang {\it et al.}~\cite{ZhichengWang12016, ZhichengWang22016} have discovered high temperature superconductivity at 29-33~K in ACa$_2$Fe$_4$As$_4$F$_2$ (A=K, Rb, Cs, 12442-family). The crystal structure is displayed in Fig. 1, where the Fe$_2$As$_2$ layers are now sandwiched between A atoms on one side and Ca$_2$F$_2$ on the other, again leading to two distinct As sites above (As$_1$) and below (As$_2$) the Fe-plane as in 1144. These materials are also situated near to optimal doping.  The electronic structure and  magnetic properties of KCa$_2$Fe$_4$As$_4$F$_2$ have been calculated based on first-principle calculations and discussed in relation to the Fe-pnictide superconductors~\cite{GuangtaoWang2016}. There are ten bands crossing the Fermi level in the nonmagnetic (NM) state, resulting in six hole-like Fermi surface (FS) sheets along the $\Gamma$-Z line and four electron-like FS sheets along the $X-P$ line. The shape of the FS is more complicated than other FeAs-based superconductors, showing multiband character. Furthermore the fixed spin moment calculations and the comparisons between total energies of different magnetic phases indicate that KCa$_2$Fe$_4$As$_4$F$_2$ has a strong tendency towards magnetism, i.e. the stripe antiferromagnetic state. It has been found that the self-hole-doping suppresses the spin-density wave (SDW) state, inducing superconductivity in the parent compound KCa$_2$Fe$_4$As$_4$F$_2$~\cite {GuangtaoWang2016}. 

\begin{figure}[t]
\vskip -0.0 cm
\centering
\includegraphics[width = \linewidth,trim={0mm 0mm 0mm 0mm},clip]{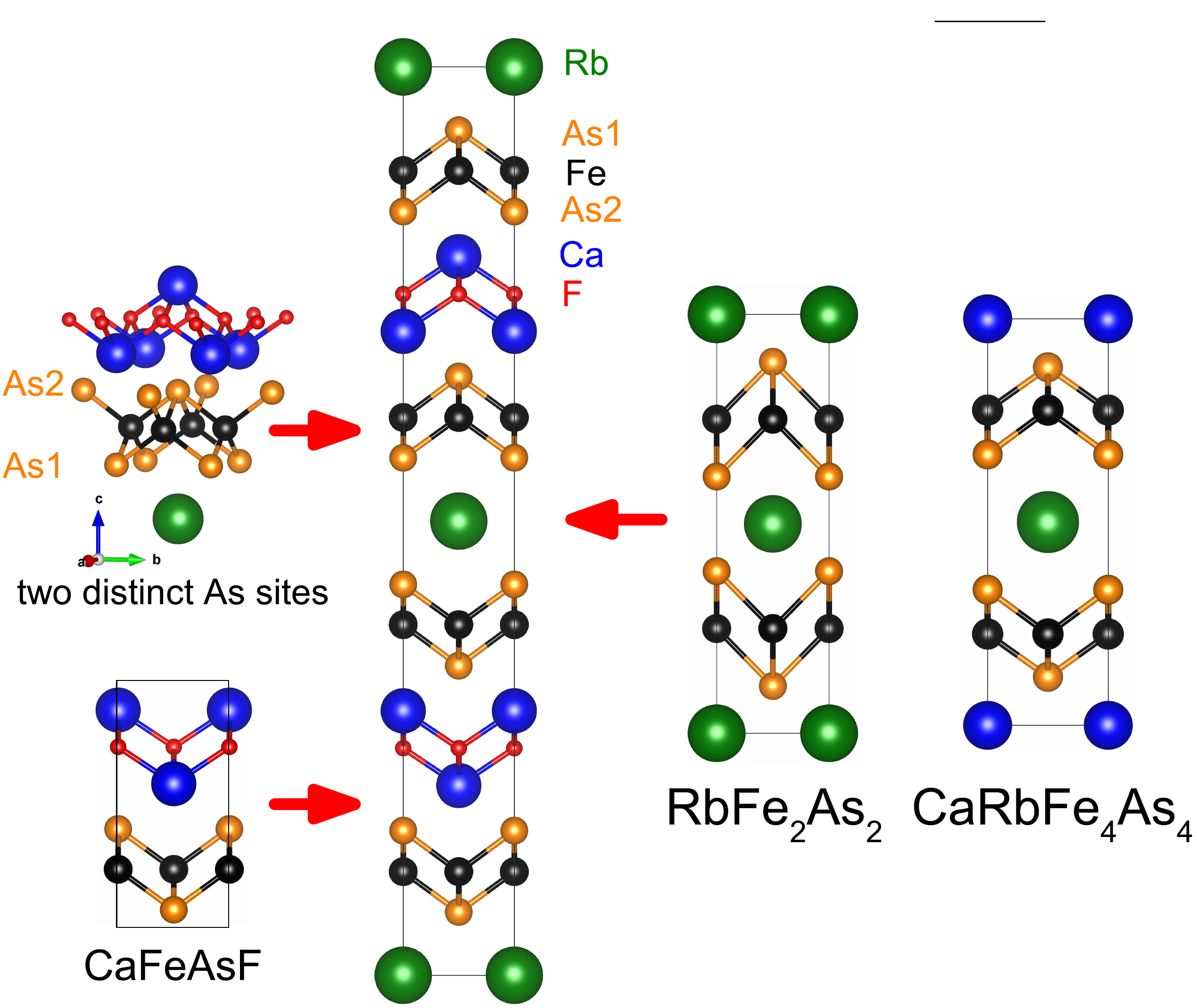}
\caption {(Color online) The tetragonal crystal structure of RbCa$_2$Fe$_4$As$_4$F$_2$. For comparison we have also given the crystal structure of RbFe$_2$As$_2$, CaRbFe$_4$As$_4$ (right side)  and CaFeAsF (left side bottom)~\cite{ZhichengWang12016, ZhichengWang22016}.}
\label{xrd:fig1}
\end{figure}

The pairing symmetry of the Cooper pairs in a superconductor is manifested in an energy gap in the single-particle excitation spectrum. The superconducting gap structure is an important characteristic for a superconductor. There is experimental evidence that cuprate-based unconventional superconductors have distinct d-wave nodal gap symmetry compared with conventional phonon-mediated superconductors which have nodeless s-wave gap~\cite{Anderson2013, Keimer2015}. On the other hand the superconducting gap symmetry in iron-based superconductors is rather more diverse and the subject of ongoing debate~\cite{122family2, JPaglione2010, H.Hosono2015, YunkyuBang2017}. Whereas nodeless gap structures have been observed in some of the doped 122-family~\cite {122family1, 122family2, JPaglione2010,  XianhuiChen2014, MRotter2008}, 1144-family~\cite{CaKFe4As4ARPES, 1144Gap1,1144Gap2,1144INS, 1144NMR}, A$_x$Fe$_2$Se$_2$ (A=K, Cs) ~\cite{Y.Zhang2011} and FeTe$_{1-x}$Se$_x$~\cite {H.Miao2012},  the signatures of nodal superconducting gaps have been reported in LaOFeP~\cite{J.D.Fletcher2009}, LiFeP~\cite{J.S.Kim2013}, KFe$_2$As$_2$~\cite{K.Hashimoto2013, J.K.Dong2010} , BaFe$_2$(As$_{1-x}$P$_x$)$_2$~\cite{Y.Zhang2012}, BaFe$_{2-x}$Ru$_x$As$_2$~\cite{xQiu2011} and FeSe~\cite{C.L.Song2010}. Furthermore  applied pressure and doping or chemical pressure change the gap symmetry from nodeless to nodal in  Ba$_{0.65}$Rb$_{0.35}$Fe$_2$As$_2$~\cite{Z.Guguchia2015} and in BaFe$_{2-x}$Ni$_x$As$_2$~\cite{MahmoudAbdel-Hafiez2015}. More interestingly the single crystal $\mu$SR study on FeSe reveals a nodeless gap (anisotropic-$s$-wave) along the c-axis, but one nodal and one isotropic ($s$+$d$-wave ) gap in the ab-plane~\cite {FeSePabi}.  \\
\indent To understand the mechanism of unconventional superconductivity and develop realistic theoretical models of Fe-based superconductors it is very important to study the pairing symmetry and the nature of the superconducting gap. There is no general consensus on the nature of pairing in iron-based superconductors leading to a variety of possibilities ranging from $s_{++}$ wave to $s_\pm$, to d wave. Furthermore it is also important to investigate whether time-reversal symmetry (TRS) in the superconducting state is preserved or not as well as the role of spin-fluctuations. Broken symmetry can modify the physics of a system and nature of the pairing, thereby resulting in novel and uncommon behavior. Muon-spin rotation and relaxation ($\mu$SR) is an ideal and sensitive microscopic  technique to investigate the properties of the superconducting state. Transverse field (TF) $\mu$SR provides information on the field distribution in the superconducting state and hence  information on the penetration depth and gap symmetry.  On the other hand zero-field (ZF) $\mu$SR allows the detection of very small internal fields and hence can provide direct information about whether TRS is preserved. Recently we have investigated the nature of the superconducting gap and TRS  in  ACa$_2$Fe$_4$As$_4$F$_2$ (A=K and Cs) compounds using $\mu$SR measurements~\cite{M.Smidman2017, F. K. K. Kirschner2017}. We found two superconducting gaps with at least one nodal gap in these compounds, but no clear sign of TRS breaking. It is therefore important to investigate the gap symmetry and TRS in A=Rb compound. Here we report TF- and ZF-$\mu$SR measurements of the A=Rb compound. Our study shows that the superfluid density derived from the depolarization rate of the TF-$\mu$SR fits better to two isotropic gaps following a $s$+$s$-wave model and ZF-$\mu$SR does not reveal any clear sign of TRS breaking below $T_c$. 

\section{Experimental Details}

The sample was characterized using powder x-ray diffraction (XRD),  magnetic susceptibility and heat capacity measurements. 
The heat capacity was measured using a Quantum Design Physical Property Measurement System (PPMS) between 1.8 and 80 K.  A standard thermal relaxation method was used with a sample mass of 8 mg. The DC magnetization measurements were carried out using a Quantum Design Magnetic Property Measurement System (MPMS). Muon spin relaxation/rotation ($\mu$SR) experiments were carried out on the MuSR spectrometer at the ISIS pulsed muon source of the Rutherford Appleton Laboratory, UK~\cite{sll}. The $\mu$SR measurements were performed in  transverse$-$field (TF), zero-field (ZF) and longitudinal field modes. A powder sample of RbCa$_2$Fe$_4$As$_4$F$_2$ was mounted on a silver  (99.999\%) sample holder.  The sample was cooled under He-exchange gas in a He-4 cryostat operating in the temperature range of 1.5 K$-$300 K.  TF$-\mu$SR experiments were performed in the superconducting mixed state in an applied field of 40 mT, well above the lower critical field of $\mu_0$$H_{c1}$ $\sim$ 20 mT (see Fig.2c) of this material.~Data were collected in the field$-$cooled (FC) mode, where the magnetic field was applied above the superconducting transition temperature and the sample was then cooled down to base temperature. Muon spin rotation and relaxation is a dynamic method that allows one to study the nature of the pairing symmetry in superconductors~\cite{js, amato}. The vortex state in the case of type-II superconductors gives rise to a spatial distribution of local magnetic fields; which demonstrates itself in the $\mu$SR signal through a relaxation of the muon polarization. Zero-field (ZF) $\mu$SR measurements were performed from 1.2~K to 150~K in the longitudinal geometry. We also performed  longitudinal field $\mu$SR measurements at 1.2~K and 35~K. The $\mu$SR data were analyzed using WiMDA~\cite{FPW}.

\begin{figure}[t]
\vskip -0.0 cm
\includegraphics[width = \linewidth, trim={0mm 0mm 0mm 34mm},clip]{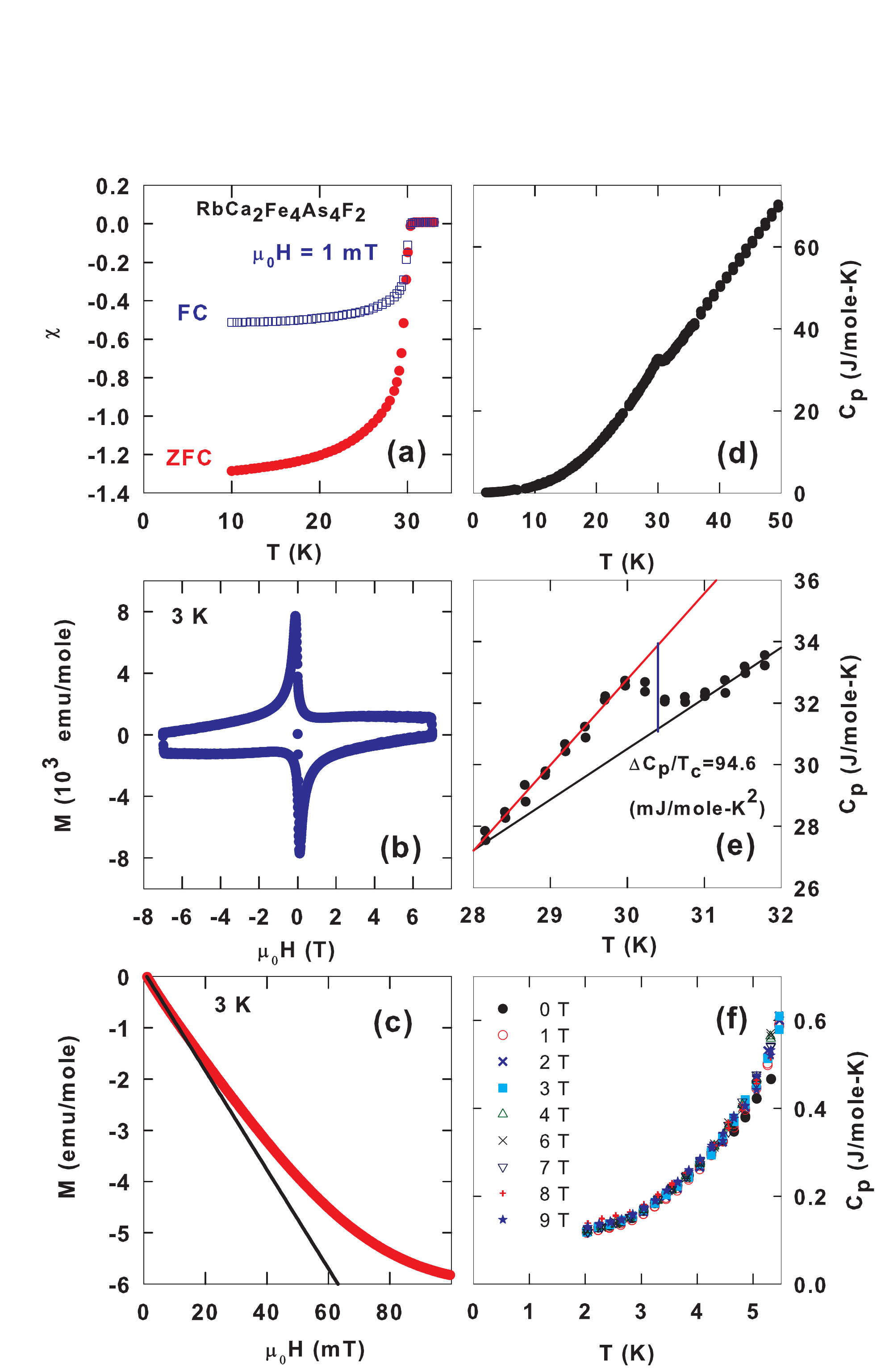}

\caption {(Color online) 
(a) Low-field dc-magnetic susceptibility measured in zero-field cooled (ZFC) and field cooled (FC) modes in an applied field of 1 mT. (b) The isothermal field dependence of magnetization at 3 K. (c) The isothermal field dependence of the magnetization at low fields at 3 K. The solid line 
shows a linear fit to the low field data.  (d) Temperature dependence of heat capacity ($C_p$) versus temperature in zero field. (e)  $C_p$ vs $T$ in an expanded scale near $T_c$. The solid lines show the linear fit above and below $T_c$ and the vertical line shows the jump in the heat capacity at  $T_c$. (f) ($C_p$) versus temperature in various applied magnetic fields up to 9~T.}
\label{bulk:fig2}
\end{figure}

\section{Results and discussions}

The analysis of the powder x-ray diffraction at 300 K reveals that the sample is single phase and crystallizes in the tetragonal crystal structure with space group $I4/mmm$ (No. 139, Z = 2) as shown in Fig. 1~\cite {ZhichengWang12016, ZhichengWang22016}. The refined values of the lattice parameters are   a = 3.8716(1)~\AA~and c = 31.667(1)~\AA. 

\par
 The low-field magnetic susceptibility measured in an applied field of 1 mT shows an onset of diamagnetism below 30 K indicating that superconductivity occurs at 30 K and the superconducting volume fraction is close to 100\% at 10~K  [Fig.~2(a)]. This result confirms the bulk nature of superconductivity with $T_c$ = 30 K in RbCa$_2$Fe$_4$As$_4$F$_2$, which is comparable to $T_c$ = 33.3 ~K and 29~K observed in ACa$_2$Fe$_4$As$_4$F$_2$ (A=K and Cs), respectively~\cite {ZhichengWang12016, ZhichengWang22016}. 

\par

The magnetization isotherm $M\left(H\right)$ curve at 3 K [Fig. 2(b)] shows typical behaviour for type-II superconductivity. The lower critical field $H_{c1}$ obtained from the M vs H plot at 3 K by linear fitting the data between 0 and 15 mT is about 20 mT [Fig.~2(c)]. The upper critical field ($\mu_0$$H_{c2}$)  measurements using the field dependent resistivity reveals the slope $d\mu_0H_{c2}$/dT$\sim$-13.9 T/K at $T_c$ ~\cite{ZhichengWang22016} and the  Pauli limit is $\mu_{0}H_{P} = 1.84T_{\mathrm{c}}$ =  55.2 T~\cite{CAM}.  Further using the orbital limiting upper critical field, $\mu_0$$H_{c2}$(0)=0.73($dH_{c2}$/dT)$_{Tc}$$T_c$, we have estimated  $\mu_0$$H_{c2}$(0)=0.30 kT. This value of $H_{c2}$ gives the coherence length $\xi$=($\Phi_0$/(2$\pi$$H_{c2}$))$^{1/2}$=1.04 nm, where $\Phi_0$= 2.07x10$^{-15}$ Tm$^2$ is the magnetic flux quantum. The specific heat $(C_{p})$ is displayed in Fig.~2(d) for zero field and an applied fields up to 9~T(Fig.2(f)). A clear anomaly is observed in the zero field $C_{p}$ corresponding to the superconducting transition at around 30.4(4)~K. The jump in  $C_{p}$  was estimated by linearly extrapolating the data above and below T$_c$, yielding a jump of $\Delta$C$_{p}$/$T_{\mathrm{c}}$ = 94.6 (mJ/mol K$^2$), which  is smaller than 150  (mJ/mol K$^2$) observed in  KCa$_2$Fe$_4$As$_4$F$_2$~\cite {ZhichengWang12016}. To shed light on the nature of the gap symmetry we also performed field dependent heat capacity measurements up to a field of 9 T. We found that the heat capacity is almost independent of applied field between 2~K and 5~K.

\begin{figure}[t]
\centering
\includegraphics[width = \linewidth]{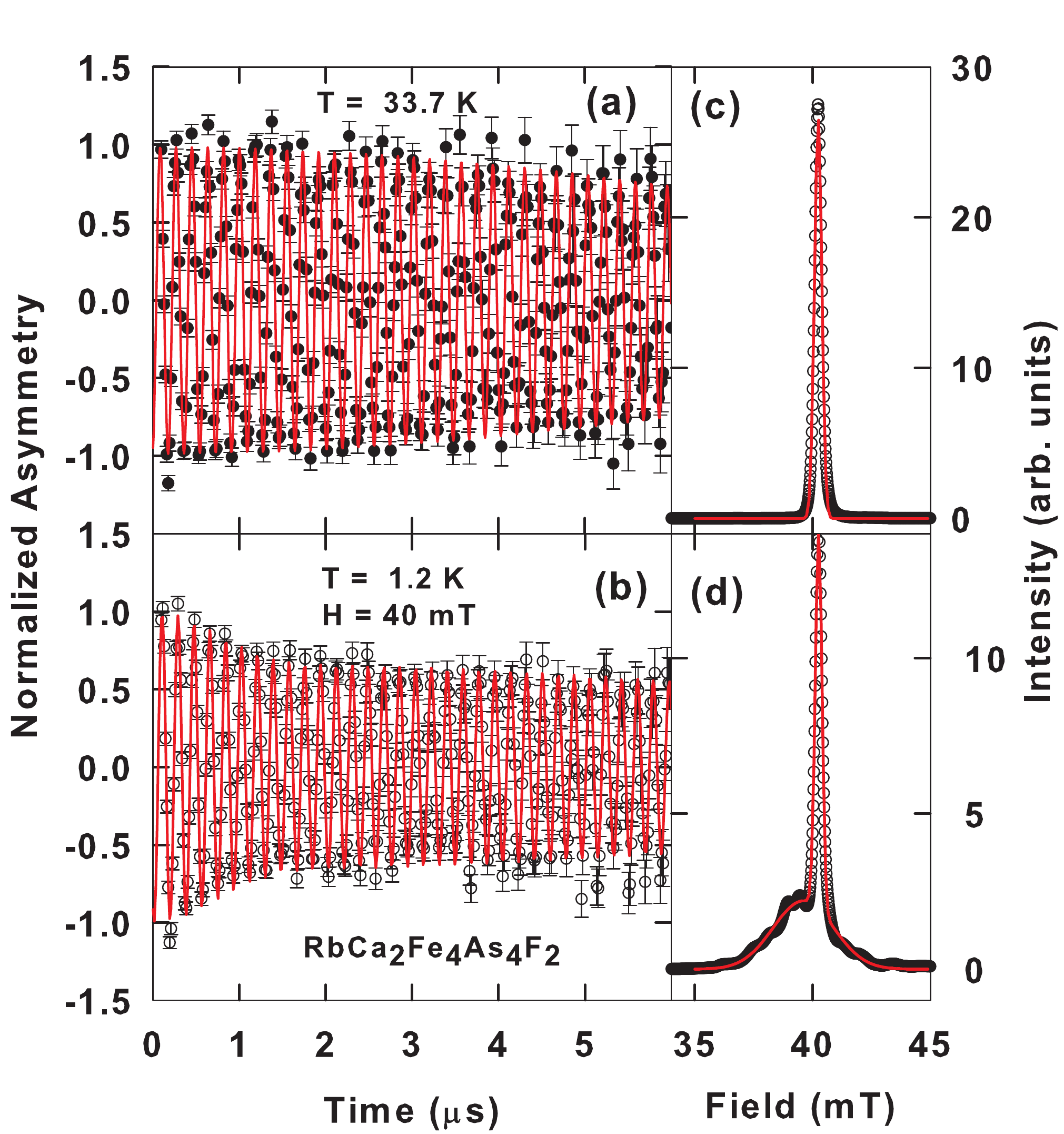}
\caption {(Color online)Muon spin rotation ($\mu$SR) measurements of RbCa$_2$Fe$_4$As$_4$F$_2$ in a transverse field of 40 mT at (a) 33.7~K (above $T_c$) and (b) 1.2~K (below $T_c$). The solid line shows a fit using Eq.(1). (c) and (d) display the corresponding maximum entropy spectra (above and below $T_c$) and the red lines show fits using one (c) and two (d) Gaussian functions.}
\label{musr1:fig3}
\end{figure} 

\par

Figures 3 (a) and (b) show the TF$-\mu$SR precession signals above and below $T_{\bf c}$ obtained in FC mode with an applied field of 40~mT (well above $H_{c1}\sim$ 20~mT but below $H_{c2}$ $>>$ 7 T, at 3 K, see Fig.2b) and Figs. 3 (c) and (d) show the corresponding maximum entropy plots, respectively.  It is clear that above $T_c$ the  $\mu$SR spectra show a very small relaxation mainly from the quasi-static nuclear moments,  and the internal field distribution is very sharp and centered near the applied field.  However at 1.2~K $\mu$SR spectra show strong damping and the internal field distribution has two components, one very sharp near the applied field and one very broad which is shifted lower than applied field.  The observed decay of the $\mu$SR signal with time below $T_{\bf c}$ is due to the inhomogeneous field distribution of the flux-line lattice. We attribute the narrow component at the applied field to muons stopping in the silver sample holder, which indicates that the field distribution within the vortex lattice is described well by one Gaussian centered at a field below 40 mT. We have used an oscillatory decaying Gaussian function to fit the TF$-\mu$SR time dependent asymmetry spectra:

\begin{equation}
A(t)=A_1{\rm e}^{-\sigma^2t^2/2}{\rm cos}(\gamma_{\mu}B_1t + \phi)+ A_2{\rm cos}(\gamma_{\mu}B_2t + \phi), 
\label{TFFit}
\end{equation}

\noindent where $\gamma_{\mu}/2\pi=135.5$~MHz/T is the muon gyromagnetic ratio, $\sigma$ is the Gaussian relaxation rate, $\phi$ is the phase, which is related to the detector geometry, $A_1$ and $A_2$ are the magnitudes of the terms from the sample and silver holder respectively, while $B_1$ and $B_2$ are respective internal fields. We grouped all detectors in 8 groups and all the groups were fitted simultaneously using the WIMDA software. The total amplitudes for each group of detectors were fixed. Furthermore, we first estimated the value of $A_1\approx$0.7 and $A_2\approx$0.3 by fitting the 1.2~K data and kept them fixed during the analysis allowing us to extract the temperature dependence of the relaxaton rate $\sigma(T)$.  Equation 1 contains the total relaxation rate $\sigma$ from the superconducting fraction of the sample; there are contributions from the vortex lattice ($\sigma_{\rm sc}$) and nuclear dipole moments   ($\sigma_{\rm nm}$) (see Fig.4b inset), where the latter is assumed to be constant over the entire temperature range  [where $\sigma$ = $\sqrt{(\sigma_{\rm sc}^2+\sigma_{\rm nm}^2)}$]. The contribution from the vortex lattice, $\sigma_{\rm sc}$, was determined by quadratically subtracting the background nuclear dipolar relaxation rate ($\sigma_{\rm nm}$=0.138(5)$\mu$s$^{-1}$) obtained from the spectra measured above $\it {T}_{\bf c}$. As the applied field (40~mT) is much less than the upper critical field ($\mu_0$$H_{c2}$ $>$ 7~T), $\sigma_{\rm sc}$ can be directly related to the effective penetration depth $\lambda_{\rm eff}$ using the following equation~\cite{Brandt2003}:

\begin{equation}
\sigma_{\rm sc}/\gamma_{\mu} = 0.0609\Phi_0/\lambda_{\rm eff}^2,
\end{equation}
where $\Phi_0$ is the magnetic flux quantum. This relation between $\sigma_{\rm sc}$ and $\lambda_{\rm eff}$ is valid for 0.13/$\kappa^{2}$$<<$(H/H$_{c2}$)$<<$1, where $\kappa$=$\lambda$/$\xi$$\gg$70~\cite{Brandt2003}. Since RbCa$_2$Fe$_4$As$_4$F$_2$ has a two-dimensional layered crystal structure with large separation between Fe$_2$As$_2$-layers, the out of plane penetration depth ($\lambda_{\rm c}$) is much larger than that in the plane ($\lambda_{\rm ab}$), so that the effective penetration depth can be estimated as  $\lambda_{\rm eff}=3^{\frac{1}{4}}\lambda_{\rm ab}$ \cite{FESENKO1991551}. 

Furthermore the penetration depth is directly related to the normalized superfluid density, n$_{\rm ns}$. In our analysis we modelled the temperature dependent normalized superfluid density using the following equation~\cite{Prozorov}

\begin{equation}
n_{\rm ns}(T)=\frac{\lambda_{\rm ab}^{-2}(T,\Delta)}{\lambda_{\rm ab}^{-2}(0)} = 1 + \frac{1}{\pi} \int_{0}^{2\pi} \int_{\Delta(T, \varphi)}^{\infty}\frac{\partial f}{\partial E}\frac{E{\rm d}E{\rm d}\phi}{\sqrt{E^2-\Delta^2(T, \varphi)}},
\label{RhoS}
\end{equation}

\noindent where $f=\left[1+\exp\left(-E/k_{\mathrm{B}}T\right)\right]^{-1}$ is the Fermi function. The temperature and angular dependence of the gap is given by $\Delta(T, \varphi)$=$\Delta_0 \delta(T/\it {T}_c)g(\varphi)$, whereas $g(\varphi)$ refers to the angular dependence of the superconducting gap function and $\varphi$ is the azimuthal angle along the Fermi surface.   We have used the BCS  formula for the temperature dependence  of the gap, which is given by $\delta(T/T_c)$ =tanh$[(1.82){(1.018(\it {T}_c/T-1))}^{0.51}]$~\cite{UBe131}. $g(\varphi)$~\cite{Annett, Pang} is given by (a) 1 for $s-$wave gap [also for $s+s$ wave gap], (b) $|$cos(2$\varphi$)$|$  for an $d-$wave gap with line nodes~\cite{Prozorov,UBe131, chia}. For the two-gap analysis, we have used a weighted sum of the two components of the resulting normalized superfluid density:
\begin{equation}
n_{\rm ns}=wn_{\rm ns}(\Delta_1, T)+(1-w)n_{\rm ns}(\Delta_2, T)
\end{equation}

\begin{figure}[t]
\vskip -1.0 cm
\centering
\includegraphics[width = \linewidth, trim={0mm 0mm 0mm 34mm},clip]{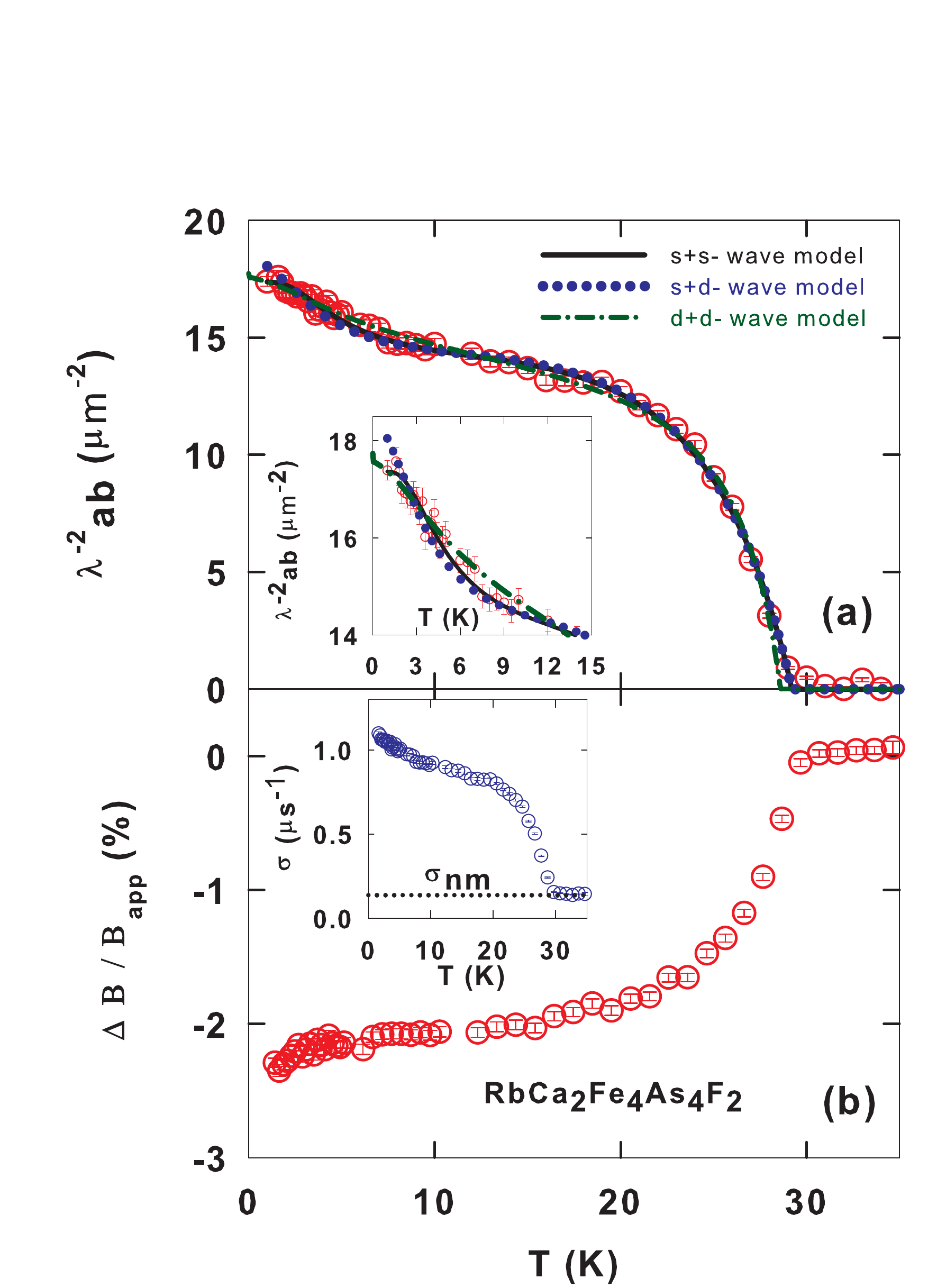}
\caption {(Color online) (a) Temperature dependence of  $\lambda_{\rm ab}^{-2}$ of RbCa$_2$Fe$_4$As$_4$F$_2$. $\lambda_{\rm ab}^{-2}$ of FC mode (symbols), where the  lines are the fits to the data using  Eq. 3 for various two-gap models. The solid black line shows the fit using an isotropic $s+s$-wave model with $\Delta_1(0)$ = 8.15$\pm0.01$ meV and $\Delta_2(0)$ = 0.88$\pm0.01$ meV, the dotted blue line shows the fit to an $s$+$d$-wave model  with  $\Delta_1(0) = 8.08\pm 0.02$ meV and  $\Delta_2(0) = 0.92\pm 0.01$ meV and the dashed-dotted green line shows the fit to a $d$+$d$-wave model with  $\Delta_1(0) = 14.05\pm 0.26$ meV and  $\Delta_2(0) = 1.26\pm 0.02$ meV. The inset shows low temperature data in an expanded scale. (b)  The normalized internal field shift as a function of temperature. The inset shows temperature dependence of total relaxation rate $\sigma$ and the dotted line shows the temperature independent contribution of nuclear depolarization rate $\sigma_{\rm nm}$.}
\label{musr2:fig4}
\end{figure}

\begin{figure}[t]
\vskip -1.0 cm
\centering
\includegraphics[width = \linewidth]{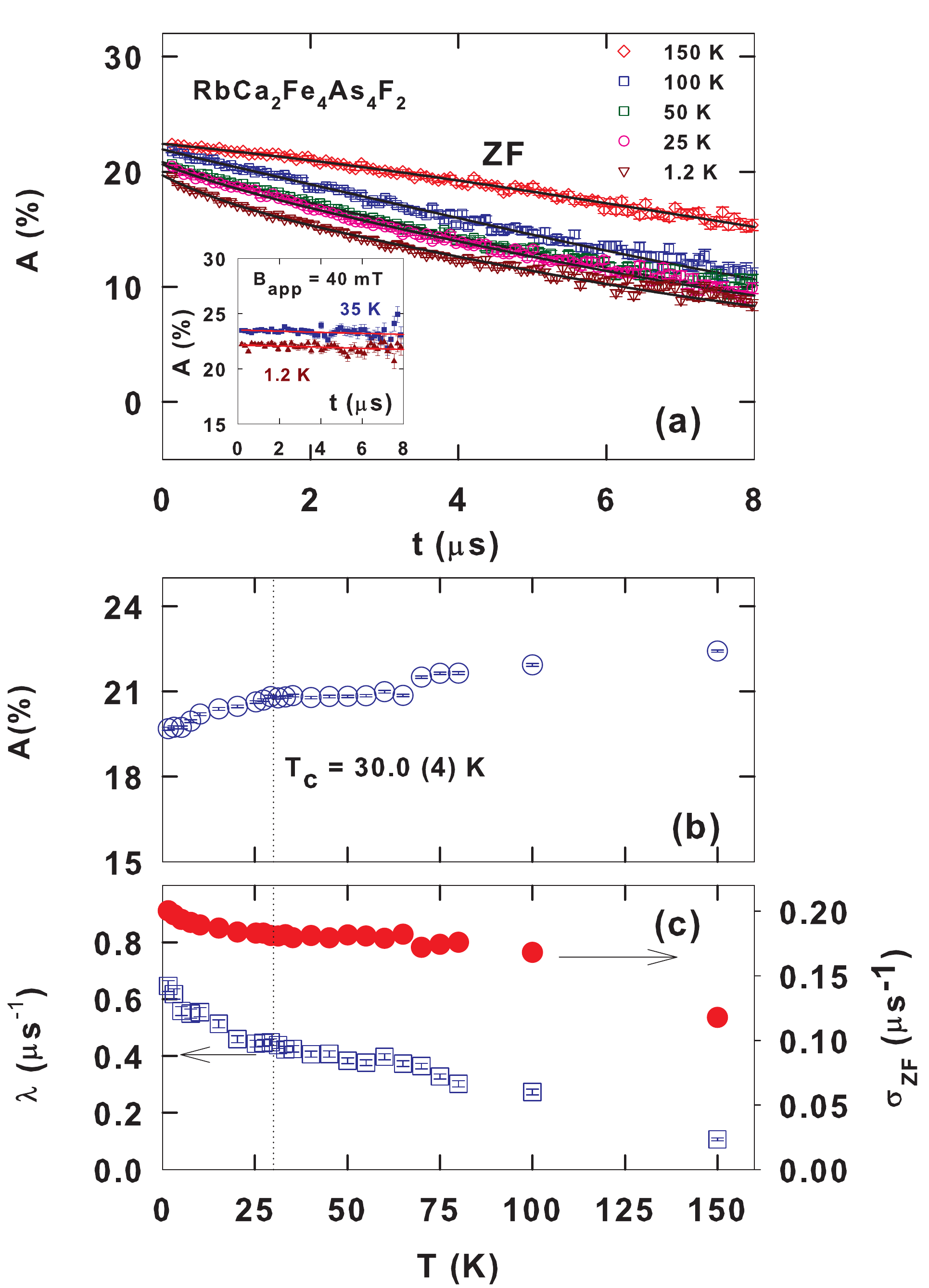}
\caption {(Color online) (a) Zero-field $\mu$SR spectra at four selected temperatures. The solid red lines show the fit described in the text. (b)-(c) The fit parameters versus temperature of zero-field $\mu$SR spectra of RbCa$_2$Fe$_4$As$_4$F$_2$. The dotted vertical line shows the transition temperature. The inset in (a) shows the spectra measured in an applied longitudinal field of 40~mT.}
\label{musr3:fig5Rb}
\end{figure}

\begin{figure}[t]
\vskip -1.0 cm
\centering
\includegraphics[width = \linewidth, trim={0mm 0mm 0mm -12mm},clip]{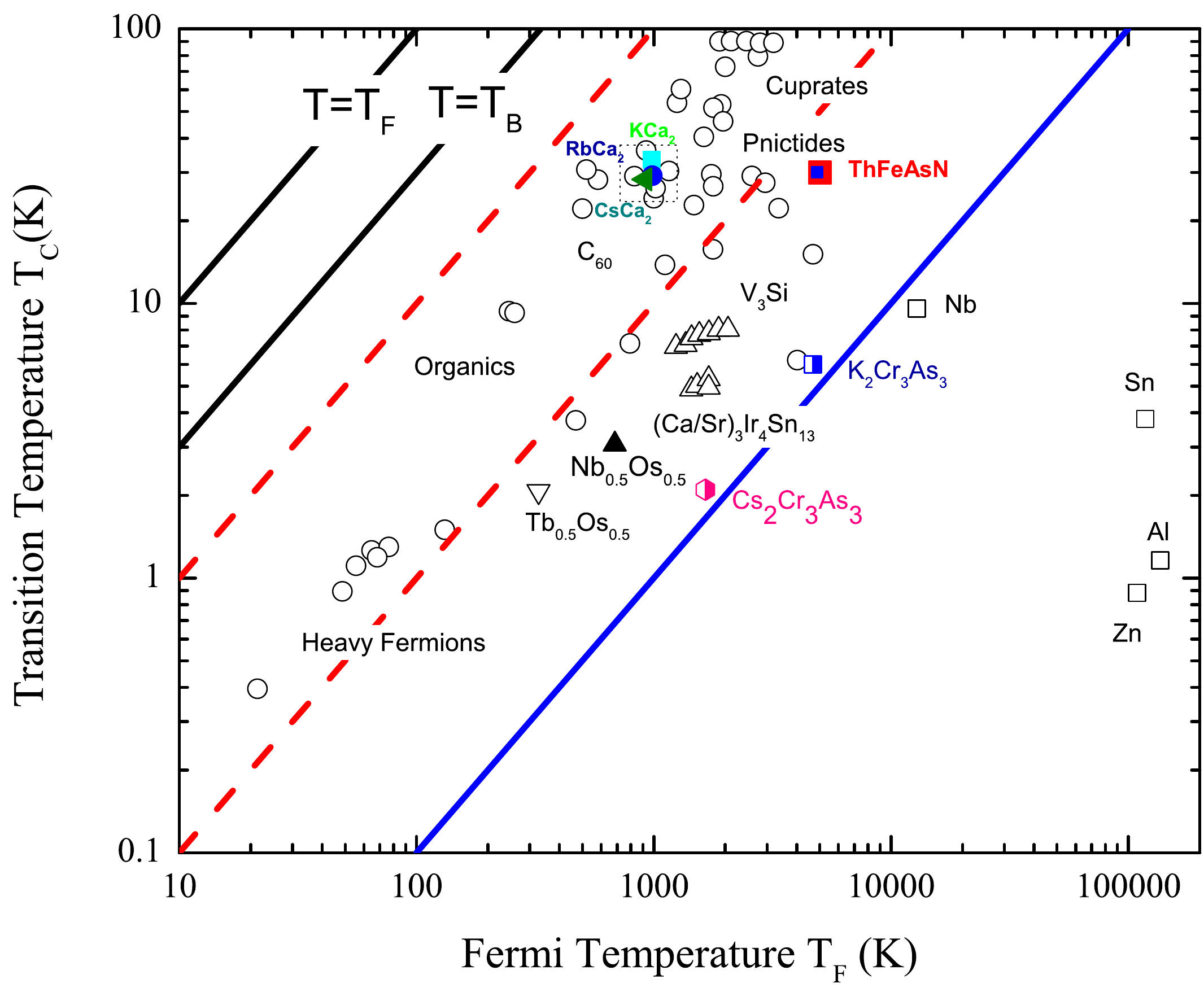}
\caption {(Color online) A schematic representation of the Uemura plot of superconducting transition temperature $T_c$  against effective Fermi temperature $T_F$. The dotted squares shows the region where ACa$_4$Fe$_2$As$_4$F$_2$ (A=K, Rb and Cs) compounds are located. The solid cyan square, solid blue circle and  dark green horizontal triangle show the points for A=K, Rb and Cs, respectively. The ``exotic'' superconductors fall within a common band for which 1/100$<T_c$/$T_F<$1/10, indicated by the region between two red color dashed lines in the figure. The solid black line correspond to the Bose-Einstein condensation temperature ($T_B$).~\cite{A1}. The positions of A=K, Rb and Cs on the plot indicate that these materials  belong to exotic superconductors family.}
\label{musr3:fig6Rb}
\end{figure}

\begin{figure}[t]
\centering
\includegraphics[width = \linewidth, trim={0mm 0mm 0mm 13.78mm},clip]{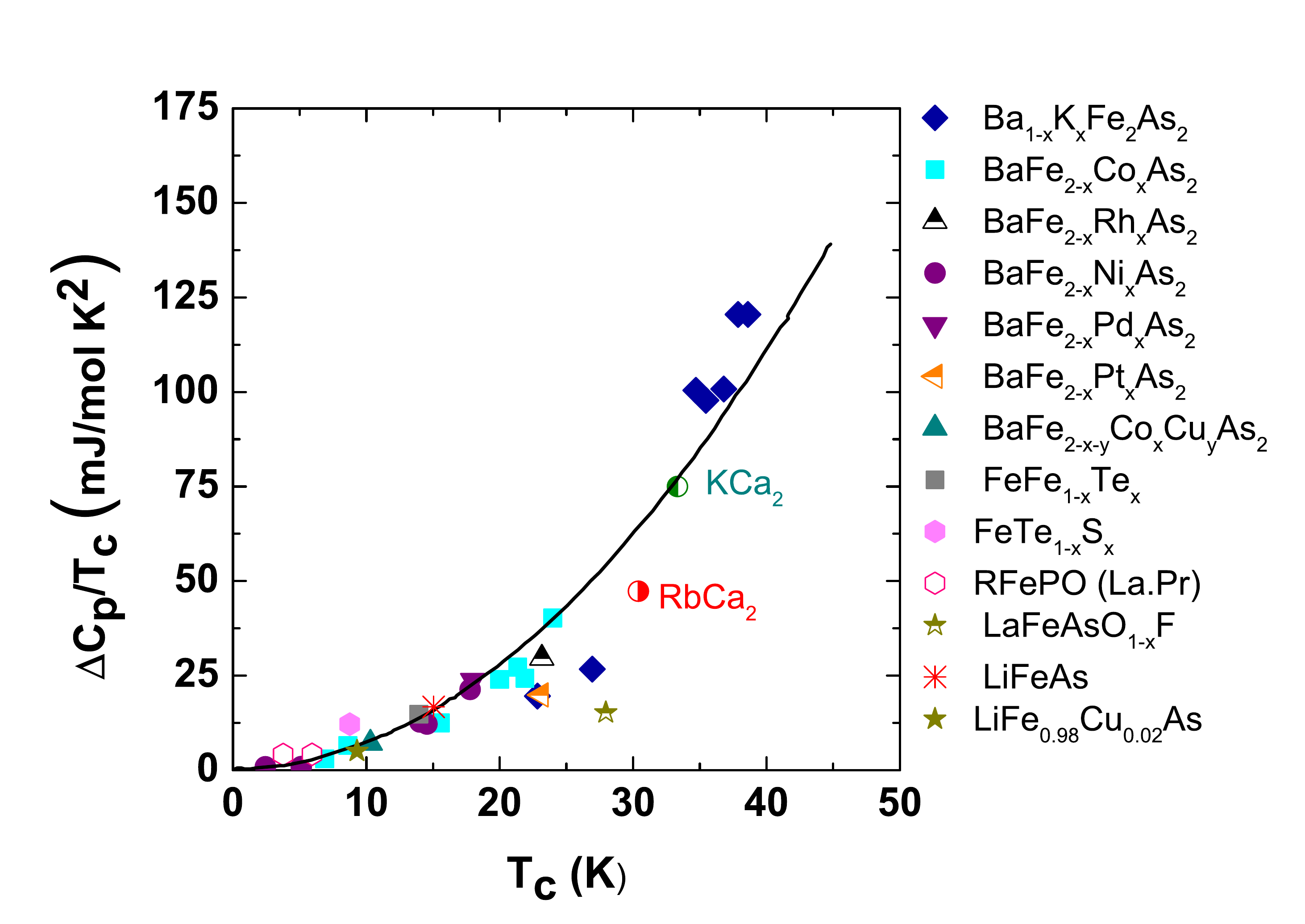}
\caption {(Color online) $\Delta$C$_p$/$T_c$ vs $T_c$ for the 122-family of FeAs-based superconductors from Ref. ~\cite{JPaglione2010, Budko2009}. The half filled circles red and blue colors (normalized by a factor 2 for comparison between 1244-family (4 Fe-atoms per formula unit) and 122-family (2 Fe-atoms per formula unit)  are for  ACa$_2$Fe$_4$As$_4$F$_2$ (A=K and Rb) compounds, respectively. The solid line is the fit to a quadratic power law ~\cite{JPaglione2010}.}
\label{fig7:fig7Rb}
\end{figure}

Figure 4 (a) shows the temperature dependence of $\lambda_{\rm ab}^{-2}$, measured in an applied field of 40 mT. $\lambda_{\rm ab}^{-2}$ increases with decreasing temperature confirming the presence of a flux-line lattice and indicates a decrease of the magnetic penetration depth with decreasing temperature. Further below 10~K $\lambda_{\rm ab}^{-2}$  shows an upturn indicating multigap behavior. The onset of diamagnetism below the superconducting transition can be seen through the decrease in the internal field below $T_c$ as shown in Fig. 4(b).  From the analysis of the observed temperature dependence of $\lambda_{\rm ab}^{-2}$, using different models for the gap, the nature of the superconducting gap can be probed. We have analyzed the temperature dependence of  $\lambda_{\rm ab}^{-2}$ based on five different models,  the single gap isotropic $s$-wave and line nodal $d$-wave models, as well as  isotropic $s$+$s$-wave, $s$+$d$-wave and $d$+$d$-wave two-gap models. It was clear from the analysis that single-gap models did not fit the data (fits are not shown). The fits to the $\lambda_{\rm ab}^{-2}$ data with various two-gap models  using Eq. (3) are shown by lines in Fig. 4(a) and the estimated fit parameters are given in Table. I. It is clear from the goodness of fitted $\chi^2$ values given in Table. I that the $d$+$d$-wave model does not fit the data very well. On the other hand the isotropic $s$+$s$-wave, $s$+$d$-wave models show good fits to the $\lambda_{\rm ab}^{-2}$ data. The value of  $\chi^2$ = 3.0 for $s$+$s$-wave model is slightly less than 3.1 for $s$+$d$-wave model. The estimated parameters for the $s$+$s$ ($s$+$d$)-wave model show one larger gap $\Delta_1(0)$ = 8.15 (8.08) (meV) and another much smaller gap $\Delta_2(0)$ = 0.88 (0.92) (meV). The smaller gap is a nodal gap in the  $s$+$d$-wave model. Our $\mu$SR analysis suggests that an $s$+$s$-wave model explains better the temperature dependence of the superfluid density than an $s$+$d$-wave model.

The value of $\lambda_{\rm ab}$(0) =  231.5$\pm$3~nm and $T_c$ = 29.19$\pm$0.04~K were estimated from the $s$+$s$-wave fit. The estimated value of 2$\Delta_1(0)$/$k_B$$\it{T}_{\bf c}$ = 6.48 from the $s$+$s$-wave fit is larger than the value 3.53 expected for BCS superconductors~\cite{B.C.S.}, indicating the presence of strong coupling and unconventional superconductivity in RbCa$_2$Fe$_4$As$_4$F$_2$. On the other hand for the smaller gap the value 2$\Delta_1(0)$/$k_B$$\it{T}_{\bf c}$ = 0.7 is much smaller than the BCS value. The two-gap nature, one larger and another smaller than the BCS value, are commonly observed in Fe-based superconductors~\cite{twogaps, twogaps1} as well as in Bi$_{4}$O$_{4}$S$_{3}$~\cite{Biswas2013}.  The observation of two isotropic gaps and nodeless superconductivity in  RbCa$_2$Fe$_4$As$_4$F$_2$ is very similar to that observed in CaKFe$_4$As$_4$, where clear evidence is found for multigap nodeless superconductivity with an $s_{\pm}$ pairing state \cite{CaKFe4As4ARPES,1144Gap1,1144Gap2,1144INS,1144NMR}. Recently we have observed two gaps in ACa$_2$Fe$_4$As$_4$F$_2$ (A=K and Cs)~\cite {M.Smidman2017, F. K. K. Kirschner2017} and ThFeAsN~\cite{ThFeAsN}, but at least one gap appears to be nodal in these compounds. Two  superconducting gaps (one larger and another smaller) were also observed in SrFe$_{1.85}$Co$_{0.15}$As$_2$,  with $T_c$ = 19.2 K in an STM study~\cite{STMtwogaps}. Moreover combined ARPES and $\mu$SR studies on Ba$_{1-x}$K$_{x}$Fe$_{2}$As$_{2}$ with $T_c$ = 32.0 K also revealed the presence of two gaps ($\Delta$$_1$ = 9.1 meV and $\Delta$$_2$ = 1.5 meV)~\cite{Khasanov2009}. The recent $\mu$SR study on FeSe single crystals revealed that the superconducting gap is most probably anisotropic $s$-wave (nodeless) along the crystallographic c-axis, but it fits better to a two-gap $s$+$d$-wave model with one nodal gap  in the ab-plane~\cite {FeSePabi}. Furthermore, nodal superconductivity has been observed in cuprate superconductors~\cite{Khasanov2007, dwave} and the recently discovered quasi-1D Cr-based superconductors, A$_2$Cr$_3$As$_3$ (A = K and Cs)~\cite{DTA1, DTA2}. 

\par

 As with other phenomenological parameters characterizing a superconducting state, the penetration depth can also be related to microscopic quantities. Within London theory~\cite{js}, $\lambda_L^2=\lambda_{\rm eff}^2= m^{*}c^2/4\pi n_s e^2$, where $m^* = (1+\lambda_{\rm e-ph})m_e$ is the effective mass and $n_s$ is the density of superconducting carriers. Within this simple picture $\lambda_L$ is independent of magnetic field. $\lambda_{\rm e-ph}$ is the electron-phonon coupling constant, which can be estimated from $\Theta_{\rm D}$ and $T_{\mathrm{c}}$ using McMillan's relation~\cite{mcm} 

\begin{equation}
\lambda_{\rm e-ph}=\frac{1.04+\mu^*\ln(\Theta_{\rm D}/1.45T_{\bf c})}{(1-0.62\mu^*)\ln(\Theta_{\rm D}/1.45T_{\bf c})+1.04},
\end{equation}
where $\mu^*$ is the repulsive screened Coulomb parameter and usually assigned as $\mu^*$ = 0.13.  As we do not have heat capacity above 80~K for the present Rb-sample, we first estimated  the value of $\Theta_{rm D}$ for KCa$_2$Fe$_4$As$_4$F$_2$ by fitting the heat capacity data between 50~K and 300~K to the Debye model ~\cite{ZhichengWang12016}, which gave  $\Theta^{\rm K}_{\rm D}$=366~K.  Then using a scaling factor~\cite{CPscaling}, which incorporates the differing molecular weight and unit-cell volume, we estimated  $\Theta^{\rm Rb}_{\rm D}$=351.6 K (similar for the Cs-sample $\Theta^{\rm Cs}_{\rm D}$=344.3 K). For RbCa$_2$Fe$_4$As$_4$F$_2$ we have used $T_{\bf c}$ = 29.19 K together with $\mu^*$ = 0.13 and have estimated $\lambda_{\rm e-ph}$ = 1.45.  This value of $\lambda_{\rm e-ph}$ is very similar to 1.38  for LiFeAs~\cite{ele-phonon_LiFeAs},  1.53 for PrFeAsO$_{0.60}$F$_{0.12}$~\cite{ele-phonon2} and 1.2 for LaO$_{0.9}$F$_{0.1}$FeAs~\cite{MUGang(2008}.  On the other hand for many Fe-based superconductors (11-family and 122-family) and HTSC cuprates (YBCO-123) smaller values of $\lambda_{\rm e-ph}$=0.02 to 0.2 and ~0.02, respectively have been reported~\cite{ele-phonon3}. Further assuming that roughly all the normal state carriers ($n_e$) contribute to the superconductivity (i.e., $n_s\approx n_e$) and using the value of $\lambda_{\rm ab}$(0) =  231.5$\pm$3~nm, we have estimated the superconducting carrier density $n_s$ and effective-mass enhancement $m^*$  to be $n_s$ = 7.45$\times$10$^{26}$ carriers/m$^3$, and $m^*$ = 2.45$m_e$, respectively.  We also estimated these parameters for ACa$_2$Fe$_4$As$_4$F$_2$ (A=K and Cs) samples (see Table-II) for comparison. 

\begin{table*}
\begin{center}
\caption{Fitted parameters obtained from the fit to the $\sigma_{\rm sc}(T)$ data of RbCa$_2$Fe$_4$As$_4$F$_2$ (as shown in Fig.~4(a)) using different gap models.\\} 
\begin{tabular}{lccccccccccccc}
\hline
\hline
 Model && T$_c$ &&Gap value &&  Gap ratio  &&   $w$ &&    {$\lambda_{\rm ab}^{-2}(0)$}       && $\chi^2$\\ 
&&K &&$\Delta_1(0)$, $\Delta_2(0)$ (meV)      &&  2$\Delta(0)/k_B T_{\bf c}$  && && $\mu$m$^{-2}$ &\\ 
\hline

$s$+$s$ wave  &&29.19(4)&& 8.15(1); 0.88(1) &&  6.48; 0.70  && 0.79(1) && 17.37(12) && 3.0\\ 
$s$+$d$ wave &&29.19(5)&& 8.08(2); 0.92(1) &&  6.42; 0.73  && 0.75(2) && 18.66(21) &&3.1\\ 
$d$+$d$ wave  &&28.57(7) && 14.05(26); 1.26(2) &&  11.41; 1.02  &&0.87(2)  && 17.82(27) && 6.2 \\ 

\hline 
\end{tabular}
\end{center}
\end{table*}
\begin{table*}
\caption{The comparison of various estimated parameters, transition temperature $T_c$, electron-phonon coupling constant, $\lambda_{\rm e-ph}$, carrier effective  mass, $m^*$, superfluid density, $n_s$, and Fermi temperature, $T_F$ of RCa$_2$Fe$_4$As$_4$F$_2$ (A=K, Rb and Cs)}
\label{Table2}
\begin{ruledtabular}
 \begin{tabular}{c c c c c c }

Compound &$T_c$(K) & $\lambda_{\rm e-ph}$ & $m^*$ ($m_e$) &$n_s$ ($10^{26}$m$^{-3}$) &$T_F$(K)  \\
\hline
\\
KCa$_2$Fe$_4$As$_4$F$_2$&33.36 &1.588&  2.588 & 8.01&741.77\\
RbCa$_2$Fe$_4$As$_4$F$_2$& 29.19 &1.451 &  2.451 &7.45&727.41\\
CsCa$_2$Fe$_4$As$_4$F$_2$ & 28.31 &1.438&  2.438 &6.66&652.32 \\

\end{tabular}
\end{ruledtabular}
\end{table*}

Zero-field $\mu$SR measurements were performed from 1.2~K to 150~K and the results are displayed in Fig.5(a) for four selected temperatures. The data were fitted with the sum of a  Lorentzian and Gaussian relaxation function 

\begin{equation}
A_0(t)=A(a{\rm exp}(-\Lambda t)+(1-a){\rm exp}(-\sigma^2_{\rm ZF}t^2/2))+A_{bg},
\end{equation}
where A$_{bg}$ is the temperature independent background arising from muons stopping on the sample  holder.  The value of A$_{bg}$=5.898(8)$\%$ and a=0.367 were estimated by fitting the 150~K data and were kept fixed during the analysis. At high temperature the relaxation is dominated by Gaussian decay, while at low temperature the relaxation changes to a Lorentzian decay. Moreover, there is a gradual decrease of initial asymmetry (A) with decreasing temperature, which suggests the development of fast component, which relaxes faster than the resolution of the experiment. The asymmetry exhibits a small drop below 70 K, which could be due to a competing magnetic/structural phase or related to some unknown phase transition and needs further investigation.  The temperature dependence of $\Lambda$ and $\sigma_{\rm ZF}$ increases with decreasing temperature between 150~K and 75~K, followed by a weak temperature  dependence between 75~K and 25~K. Below 25~K both $\Lambda$ and $\sigma_{\rm ZF}$ show a moderate temperature dependence. These results suggest the presence of weak magnetic fluctuations, but  neither quantity shows a  detectable anomaly upon passing through $T_c$, indicating an absence of time reversal symmetry breaking. However, since $\Lambda(T)$ and $\sigma_{\rm ZF}(T)$ show some temperature dependence, a weak increase of the relaxation due to time reversal symmetry breaking cannot be entirely ruled out~\cite{Re6ZrTRS, DTA2, ThFeAsNZF}. Furthermore we also performed longitudinal fields (LF) measurements at 1.2 K and 35 K in applied LF of 25, 40 and 50 mT and the data of 40 mT field are shown in the inset of Fig.~5a. At all applied longitudinal fields, the data showed negligible relaxation (i.e the asymmetry is almost constant with time), indicating very weak spin-fluctuations which require a very small LF field to decouple the $\mu$SR signal. 

\par
The correlation between $T_c$ and $\sigma_{\rm sc}(0) $ (or $\lambda_{\rm ab}^{-2}(0)$) observed in $\mu$SR studies has suggested a
new empirical framework for classifying superconducting materials~\cite{U1}. Here we explore the role of muon spin relaxation rate/penetration depth in the superconducting state for the characterisation and classification of superconducting materials as first proposed by Uemura {\it et al.}~\cite{U1}.  In particular we focus upon the Uemura classification scheme which considers the correlation between the superconducting transition temperature, $T_c$, and the effective Fermi temperature, $T_F$, determined from $\mu$SR measurements of the penetration depth~\cite{A1}. Within this scheme strongly correlated ``exotic'' superconductors, i.e. high $T_c$ cuprates, heavy fermions, Chevrel phases and the organic superconductors, form a common but distinct group characterised by a universal scaling of $T_c$ with $T_F$ such that 1/10$>(T_C/T_F)>$1/100 (Fig. 6). For conventional BCS superconductors 1/1000$>(T_c/T_F$). Considering the value of $T_c/T_F$ = 0.04 for RbCa$_2$Fe$_4$As$_4$F$_2$ (see Fig. 6), this material can be classified as an exotic superconductor, according to Uemura's classification~\cite{U1}. Furthermore we have also plotted the data of ACa$_4$Fe$_2$As$_4$F$_2$ (A=K and Cs) in Fig. 6, which also belong to the same class.
\par
It has been found that the jump in the heat capacity $\Delta$C$_p$/$ T_c$ at $T_c$ is also related to $T_c$ for electron and hole doped BaFe$_2$As$_2$ superconductors~\cite {JPaglione2010, Budko2009}. We have plotted the heat capacity jump of ACa$_2$Fe$_4$As$_4$F$_2$ (A=K and Rb) on the scaling plot shown in Fig.7. It is clear that for A=K and Rb compounds the heat capacity jump also follows this trend suggesting a common relation between $\Delta$C$_p$/$ T_c$$\sim$$T_c$$^{2}$, the so-called BNC scaling~\cite{Budko2009}.

\par

\section{Conclusions} 

In conclusion, we have presented magnetization, heat capacity and transverse field (TF) and zero-field  (ZF) muon spin rotation ($\mu$SR) measurements in the normal and  the superconducting state of RbCa$_2$Fe$_4$As$_4$F$_2$, which has a double Fe$_2$As$_2$ layered tetragonal crystal structure.  Our  magnetization and heat capacity measurements confirmed the bulk superconductivity with T$_c $ = 30.0 (4) K. From the TF $\mu$SR we have determined the muon depolarization rate in the FC mode associated with the vortex-lattice.  The temperature dependence of the superfluid density  fits  better to a two-gap model, with either an isotropic $s$+$s$-wave or an $s$+$d$-wave gap, than to single gap isotropic $s-$wave or $d$-wave models. The  $s$+$s$- and $s$+$d$-wave model fits give a goodness of fit ($\chi^2$) value of 3.0 and 3.1, respectively, suggesting than an $s$+$s$-wave model is an appropriate for the gap structure of RbCa$_2$Fe$_4$As$_4$F$_2$. Furthermore, the value (for the larger gap) of 2$\Delta_1(0)/k_{\mathrm{B}}T_{\mathrm{c}}$ = 6.48$\pm0.08$ obtained from the $s$+$s$-wave model fit is larger than 3.53, expected for BCS superconductors,  indicating the presence of strong coupling superconductivity, that is supported through a larger value of $\lambda_{\rm e-ph}$,  in RbCa$_2$Fe$_4$As$_4$F$_2$. Moreover, two superconducting gaps have also been observed in the Fe-based families of superconductors, including  in other ACa$_2$Fe$_4$As$_4$F$_2$ (A=K and Cs) compounds and hence our observation of two gaps is in agreement with the general trend observed in Fe-based superconductors. It is an open question why the A=Rb material is more consistent with two isotropic gaps, while A=K and Cs have at least one nodal gap despite the ionic size (lattice parameters and unit cell volume) increasing, while $T_c$ decreases linearly, going down the alkali atom group from K to Cs. 
Further confirmation of the presence of two gaps and their symmetry in RbCa$_2$Fe$_4$As$_4$F$_2$ could be found from angle-resolved photoemission spectroscopy (ARPES) study and TF-$\mu$SR study on single crystals, for H$\parallel$c-axis and H$\parallel$ab-plane, of RbCa$_2$Fe$_4$As$_4$F$_2$.

\section*{ACKNOWLEDGEMENT}

This work is supported by EPSRC grant EP/N023803. F.K.K.K. thanks Lincoln College, Oxford, for a doctoral studentship and National Key R and D Program of China (Grant No. 2017YFA0303100). DTA would like to thank  the Royal Society of London for the UK-China Newton mobility funding. DTA and ADH would like to thank CMPC-STFC, grant number CMPC-09108, for financial support.

\end{document}